\documentstyle[aps,epsfig,amsmath,amssymb,twocolumn]{revtex}

\def\eq#1{Eq.~(\ref{#1})}
\def\fig#1{Fig.~\ref{#1}}

\def\sec#1{Sec.~\ref{#1}}  

\begin{document}

\title{Phase behavior and material properties of hollow nanoparticles} 
\author{U. S. Schwarz and S. A. Safran}
\address{Department of Materials and Interfaces,
Weizmann Institute of Science,
Rehovot 76100, Israel}
\maketitle

\begin{abstract}
  Effective pair potentials for hollow nanoparticles like the ones
  made from carbon (fullerenes) or metal dichalcogenides (inorganic
  fullerenes) consist of a hard core repulsion and a deep, but
  short-ranged, van der Waals attraction. We investigate them for
  single- and multi-walled nanoparticles and show that in both cases,
  in the limit of large radii the interaction range scales inversely
  with the radius, $R$, while the well depth scales linearly with $R$. We
  predict the values of the radius $R$ and the wall thickness $h$ at which
  the gas-liquid coexistence disappears from the phase diagram. We
  also discuss unusual material properties of the solid, which include
  a large heat of sublimation and a small surface energy. 
\end{abstract}            

\section{Introduction}

The phase behavior and material properties of atomic as well as
colloidal systems can often be understood using the concept of an
effective pair potential \cite{a:hans86,a:russ89}.  However, in the
presence of attractive interactions there is a fundamental difference
between the atomic and the colloidal case: while atomic systems
usually have a range of attraction which is comparable with or larger
than the hard core diameter, colloidal systems have a range of
attraction that can be much smaller. One example is the attractive
depletion interaction in systems with sterically stabilized colloids
and non-adsorbing polymers.  The range of interaction can be tuned by
the polymer's radius of gyration and usually is one order of magnitude
smaller than the hard core diameter \cite{n:gast83}. Another example
is the nature of attractive surfactant interactions between inverse
microemulsion droplets. Here the range of interaction is fixed by some
microscopic interpenetration length. By tuning the droplet size, that
is the amphiphile/oil ratio, the hard core diameter can be made more
than one order of magnitude larger than the interaction range
\cite{a:huan84}.

For Lennard-Jones systems, the potential depth determines the
temperature and the hard core diameter the density scale,
respectively; after rescaling to reduced units for temperature and
density, no other degrees of freedom are left and phase transitions
fall on universal curves (\emph{law of corresponding states}). The
resulting phase diagram features a critical point for gas-liquid
coexistence and a fluid-solid coexistence which is first order at all
temperatures. For more complicated interaction potentials, the
topology of the phase diagram can change. It is long known that the
gas-liquid coexistence disappears for small interaction ranges, as
is the case for the short-ranged depletion interaction in systems
with sterically stabilized colloids and non-adsorbing polymers
\cite{n:gast83}.  Recent computational \cite{n:bolh94} and
analytical \cite{a:daan,n:teje} work suggests
that the liquid phase disappears and there is only vapor-solid
coexistence when the range of attraction is smaller than about
one-third of the hard core diameter. Moreover, an isostructural
solid-solid transition appears when the range of attraction decreases
by another order of magnitude.

Recently, the phase behavior of $C_{60}$ (buckyballs) has attracted
much attention since it constitutes a borderline case for the
disappearance of the liquid phase. The van der Waals (vdW) interaction
between buckyballs is usually modeled by the \emph{Girifalco
potential} which is obtained by integrating a Lennard-Jones potential
for carbon atoms over two spherical shells \cite{n:giri92}. It is
expected to be valid above 249 K, since then the buckyballs rotate
freely; for lower temperatures, the interaction becomes anisotropic
\cite{n:hein91}.  Since the length scale for attraction is set by the
position of the minimum of the Lennard-Jones potential ($3.9\ \AA$),
it is smaller than the buckyballs' diameter ($7.1\ \AA$). Early
simulations using this potential gave conflicting results in regard to
the existence of a liquid phase: while molecular dynamics simulations
predicted that a liquid phase exists \cite{n:chen93}, a Gibbs ensemble
Monte Carlo simulation predicted that $C_{60}$ might be the first
non-colloidal substance found which does not have a liquid phase at
all \cite{n:hage93}. Recent Monte Carlo simulations however confirm
the existence of a small liquid phase region in the phase diagram
\cite{n:cacc97,n:hase99}.

However, carbon buckyballs are only one example of hollow nanoparticles.
Until now, more than 30 other materials which are similarly layered
have already been prepared as hollow nanoparticles (either
spherical or cylindrical), including e.g.\ the metal dichalcogenides
MeX$_2$ (Me = W, Se, X = S, Se), BN, GaAs and CdSe \cite{n:review}.
In fact, the formation of closed structures is generic for anisotropic
layered materials of finite size due to the line tension resulting from
dangling bonds. Effective pair potentials for such hollow
nanoparticles are isotropic for spherical shapes, but still depend on
their radius $R$ and thickness $h$.  In particularly, for carbon onions and
hollow metal dichalcogenides nanoparticles (inorganic fullerenes) the
thickness $h$ can vary because the particles are multi-walled.  Carbon
onions with hundreds of shells have been observed
\cite{n:ugar}.  Onions are formed by metal dichalcogenides
with up to 20 shells \cite{n:tenn92}.  In both
cases, outer radii $R$ can reach 100 nm, that is several orders of
magnitude more than buckyballs with $R = 3.55\ \AA$.  Unfortunately,
control of size and shape is still rather difficult, and only some
fullerenes can be produced in a monodisperse manner and in macroscopic
quantities.  The best investigated case are of course buckyballs,
which readily crystallize into \emph{fullerite}.  However, the
synthesis of hollow nanoparticles is a rapidly developing field, and
it is quite conceivable that more experimental systems of this kind
will be available soon, since hollow nanoparticles provide exciting
prospects for future applications in nanoelectronics and -optics, for
storage and delivery systems, for atomic force microscopy and
tribological applications \cite{n:review}.

Although this paper is concerned with hollow nanoparticles made from
anisotropic layered material, there are other types of hollow
nanoparticles which should show similar physical properties.
Colloidal core-shell particles with a porous shell can be made hollow
by removing the core through calcination or exposure to suitable
solvent.  Hollow polyelectrolyte shells have been prepared by
colloidal templating and layer-by-layer deposition \cite{n:dona98}.
The same technique can be used also to synthesize hollow silica
particles \cite{n:caru98}. Since wet chemistry and electrostatic
self-assembly allow for good control of size and shape, the wall
thicknesses $h$ could be tuned from tens to hundreds of nanometer by
varying the number of deposition cycles; the radius $R$ is determined
by the size of the templating polymer particles (usually in the
sub-micron range). Hollow nanoparticles also occur in biological
systems. Clathrin coats are hollow protein cages which induce
transport of specific membrane receptors into the cytoskeleton by
coating budding vesicles, but which also self-assemble in a test tube.
Recently, the highly regular structure of the smallest clathrin coat
(radius 35 nm) has been resolved in great detail by cryo-electron
microscopy \cite{n:smit98}.

In this work, we address the question of how the phase behaviour of
hollow nanoparticles depend on radius $R$ and wall thickness $h$. For
this purpose, we use the same framework as the work based on the
Girifalco potential, that consists of integrating a Lennard-Jones
potential for single atoms over the appropriate geometries and
investigating the phase behavior resulting from the effective
potential. We also predict how the material properties of the solid
vary with $R$ and $h$ and discuss how elastic deformations will modify
our results.

Our main result is that for both single- and multi-walled
nanoparticles, in the limit of large radii, the interaction range
scales inversely with the radius, $R$, while the well depth scales
linearly with $R$. For the phase behavior this means that the
gas-liquid coexistence will disappear with increasing $R$. Our full
analysis predicts that this will happen for single-walled
nanoparticles around $R_c = 12\ \AA$ and for completely filled
nanoparticles around $R_c = 35\ \AA$.  It follows that buckyballs,
which have $R = 3.55\ \AA$, are far from losing their coexistence
region. We find that the effect of the wall thickness $h$ is rather
small: $R_c$ initially increases with increasing $h$, but then levels
off on the atomic scale set by the Lennard-Jones potential between
single atoms. Moreover the temperature range is essentially set by $R$
and is hardly affected by $h$. All our results are derived twice, once
from the analytically accessible Derjaguin approximation and once from
a numerical treatment of the full potential.  We find that the
Derjaguin approximation predicts the right trends and provides
physical insight into the underlying mechanisms, but overestimates
both potential depth and interaction range.

We also show that crystals of hollow nanoparticles
will have unusual material properties. Their heat of sublimation
(which is already unusually high for fullerite) scales linearly with
$R$; at the same time, their surface energy (which for fullerite is
already smaller than for graphite) scales inversely with $R$. Thus
with increasing $R$ it gets more and more difficult to melt the
crystal, although it becomes more and more unstable in regard to
cleavage. The thickness $h$ does not have any significant effect here
since the particles are nearly close packed and the vdW interaction
saturates quickly over atomic distances under adhesion conditions.
However, we show that an important effect of the thickness $h$ is to
suppress elastic deformations for increasing $h$. If one wants to
exploit the unusual properites of hollow nanoparticles, multi-walled
variants are favorable as they suppress elastic effects which will
prevent gas-liquid coexistence and crystallization.

The paper is organized as follows: in
\sec{subsec:InteractionPotentialsA} we discuss the Lennard-Jones
potential for the interaction between single atoms. In
\sec{subsec:InteractionPotentialsB} we integrate it over the
geometries of single- and multi-walled nanoparticles to find effective
interaction potentials, which differ considerably from the initial
Lennard-Jones potential due to the additional length scales
introduced. Analytical predictions for the potential depth and
interaction range can be found using Derjaguin approximations, which
describe well the relevant part of the full potential in the limit of
large $R$ and which are derived in
\sec{subsec:InteractionPotentialsC}. In \sec{sec:MaterialProperties}
the Derjaguin approximations are used to predict the scaling of the
material properties of the crystal with radius $R$ and thickness $h$.
In \sec{sec:PhaseBehavior} we investigate both the full potentials and
their Derjaguin approximations to predict at which values of $R$ and
$h$ the gas-liquid coexistence will disappear from the phase diagram.
In \sec{sec:Elasticity} we discuss briefly the effect of elastic
deformations, and finally present our conclusions in
\sec{sec:Conclusion}.

\section{Effective interaction potentials}
\label{sec:InteractionPotentials}

\subsection{Lennard-Jones potential for single atoms}
\label{subsec:InteractionPotentialsA}

The calculation of vdW interactions between macroscopic regions of
condensed matter is a well-investigated subject in colloidal science
\cite{a:maha76,a:russ89,a:isra92,a:safr94}. It is well known that the
geometrical aspect of this problem is well treated by a pairwise
summation of the microscopic interaction; the strength of the
interaction (\emph{Hamaker constant}) can be derived from Lifshitz
theory or from comparision with experiment. The $1/r^6$-vdW
interaction and its crossover to Born repulsion at small separation is
usually well described by a Lennard-Jones potential:
\begin{equation}
  \label{eq:LJ12-6}
  V_{LJ}(r) = \frac{B}{r^{12}} - \frac{A}{r^6} 
= 4 \epsilon \left[ \left(\frac{\sigma}{r}\right)^{12} 
                  - \left(\frac{\sigma}{r}\right)^6 \right]\ . 
\end{equation}
Girifalco obtained the effective interaction potential between two
buckyballs by integrating \eq{eq:LJ12-6} over two spherical shells
\cite{n:giri92}. By fitting resulting predictions for energy of
sublimation and lattice constant for fullerite to the experimental
results, he found the values $A = 32 \times 10^{-60}$ erg cm$^6$ and
$B = 55.77 \times 10^{-105}$ erg cm$^{12}$ for the Lennard-Jones
interaction of carbon atoms.  Very similar values have been extracted
from an analogous procedure for graphite sheets \cite{n:giri56}.  For the
following it is useful to characterize the Lennard-Jones potential by
its hard core diameter $\sigma = (B / A)^{1/6}$ and its potential
depth $\epsilon = A^2 / 4 B$. For carbon, $\sigma = 3.47\ \AA$ and
$\epsilon = 4.59 \times 10^{-15}$ erg $= -0.11$ kT (where k is
the Bolzmann constant and T room temperature).

In order to use the atom-atom potential from \eq{eq:LJ12-6} within the
framework of a continuum approach, one has to know the density of
atoms. For single- and multi-walled nanoparticles, area density
$\varrho$ and volume density $\rho$ will be used, respectively.  The
in-plane elasticity of a carbon sheet is so strong that its in-plane
structure is hardly changed when the sheet is wrapped onto a
non-planar geometry. For example, the area density of carbon atoms in
a buckyball, $\varrho = 60 / 4 \pi R^2 = 0.38\ \AA^{-2}$ with radius
$R = 3.55\ \AA$, is very close to the one in graphite.  The area
density for both WS$_2$ and MbS$_2$ is $\varrho = 0.35\ \AA^{-2}$; it
is close to the value for carbon, since the effect of larger bond
length is offset by the existence of triple layers.  For future
purpose we define a dimensionless area density of atoms, $\tau =
\varrho \sigma^2$, and a dimensionless volume density of atoms, $\chi
= \rho \sigma^3$. The two densities are related by $\chi = \tau \sigma
/ l$, where $l$ is the distance between layers. For carbon and
inorganic fullerenes, $l = 3.4\ \AA$ and $l = 6.2\ \AA$, respectively.
For carbon one then finds $\tau = 4.56$ and $\chi = 4.60$. For the
metal dichalcogenides, $\tau$ is roughly the same, but $\chi$ is
smaller by a factor $2$ when compared with carbon.  In any case, the
dimensionless densities squared roughly amount to one order of
magnitude.

Although the molecular structure of the metal dichalcogenides is quite
different from that of carbon, the values for $\sigma$ and $\epsilon$
are expected to be similar.  From the Lennard-Jones potential given
above, the effective Hamaker constant for carbon follows as $\pi^2
\rho^2 A = 3.9 \times 10^{-12}$ erg $\approx 100$ kT. Indeed this is
the right order of magnitude for the Hamaker constant of vdW solids in
vacuum \cite{a:isra92}, thus in the following we assume that the
values of $\sigma$ and $\epsilon$ given for carbon will give the right
order of magnitude results for the metal dichalcogenides, too.

\subsection{Full interaction potentials}
\label{subsec:InteractionPotentialsB}

\begin{figure}
  \begin{center}
    \epsfig{file=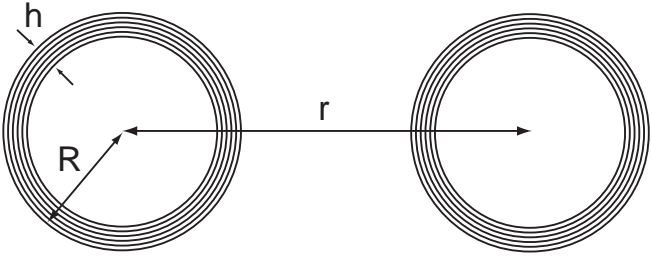,width=\columnwidth}
  \end{center}
  \caption{Schematic drawing of two multi-walled hollow nanoparticles 
    with center-to-center distance $r$, particle radius $R$ and wall
    thickness $h$.}
  \label{fig:Interaction}
\end{figure}

\begin{figure}
  \begin{center}
    \epsfig{file=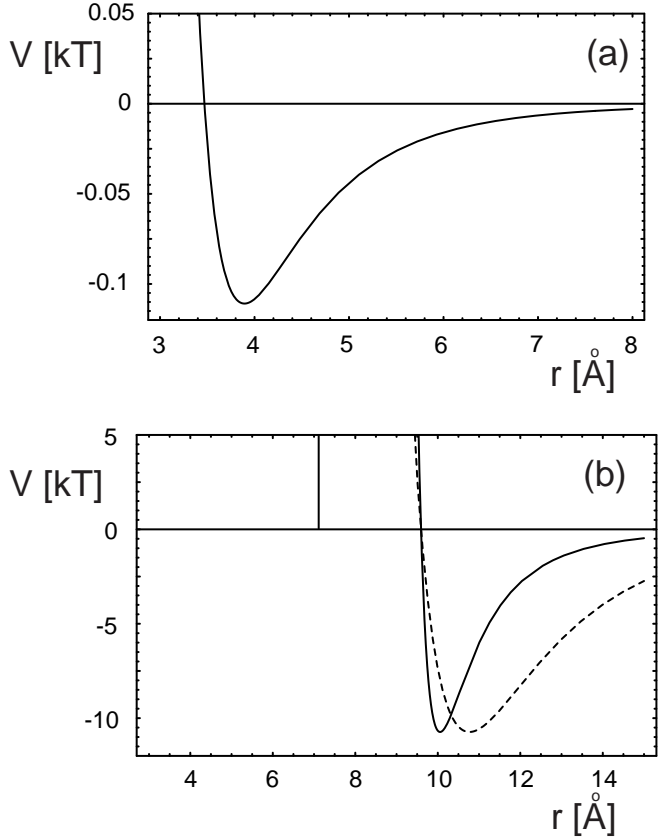,width=\columnwidth}
  \end{center}
  \caption{Different van der Waals interaction potentials: the Lennard-Jones 
    potential for carbon atoms (a) is two orders of magnitude
    weaker than the Girifalco potential for buckyballs (b, solid).
    A Lennard-Jones potential with the same zero-crossing and
    potential depth as the Girifalco potential (b, dashed) shows
    that the Girifalco potential is rather short-ranged. The vertical
    line indicates that the potential diverges at a finite value of
    separation.}
  \label{fig:Girifalco}
\end{figure}

When integrating \eq{eq:LJ12-6} over the volumes of two hollow
nanoparticles, we introduce two additional length scales: the radius
$R$ and the thickness $h$. \fig{fig:Interaction} depicts the geometry
of two multi-walled hollow nanoparticles.  For the following it is
useful to define the two dimensionless quantities $\eta = 2 R /
\sigma$ and $\nu = h / \sigma$, that are the particles' diameter and
thickness, respectively, in units of the Lennard-Jones hard core
diameter. Since $h \le R$, we have $2 \nu \le \eta$.  The Girifalco
potential for two single-walled nanoparticles of radius $R$ separated
by a distance $r$ follows from integrating \eq{eq:LJ12-6} over two
spherical shells \cite{n:giri92}.  Using the definitions given above,
it can be written as
\begin{align} 
\label{eq:Girifalco}
V_G(r) & = \pi^2 \tau^2 \epsilon \Bigg[ 
\frac{2}{45 \eta^8} \left( \frac{1}{s (s-1)^9} 
+ \frac{1}{s (s+1)^9} - \frac{2}{s^{10}} \right) \nonumber \\
& - \frac{1}{3 \eta^2} \left( \frac{1}{s (s-1)^3} 
+ \frac{1}{s (s+1)^3} - \frac{2}{s^{4}} \right) \Bigg]
\end{align}
where $s = r / 2 R$ is the distance in units of the particles'
diameter. Here $\nu = h / \sigma$ does not appear since the shell is
assumed to be infinitely thin.  For buckyballs, $\eta = 2 R / \sigma =
2.05$. Then the potential minimum is at $r_0 = 10.06\ \AA$ with a
potential depth of $V_0 = - 4.44 \times 10^{-13}$ erg $= -10.73$ kT.
In \fig{fig:Girifalco}a we plot the Lennard-Jones potential from
\eq{eq:LJ12-6} for carbon atoms and in \fig{fig:Girifalco}b the
Girifalco potential from \eq{eq:Girifalco} for buckyballs. The
integration over the given geometry has strongly changed the character
of the interaction potential: it is now two orders of magnitude
stronger and rather short ranged. The concept of a small interaction
range will be quantified below. The small width of the potential well
becomes evident when compared with a Lennard-Jones potential with same
effective diameter and potential depth (dashed line in
\fig{fig:Girifalco}b).  The geometrical effect on the effective
interaction potential is a subject well-known from colloid science.

We now calculate the vdW interaction between multi-walled hollow
nanoparticles, that is between two thick shells. We consider two balls
$B_i$ ($i = 1,2$), each of which consists of a thick shell $S_i$ and a
core $C_i$, which is a ball with smaller radius.  Then the interaction
between the two shells can be expressed in terms of interactions
between several balls:
\begin{align}
  V_{S_1, S_2} & = V_{B_1, B_2} - V_{C_1, C_2} -  V_{C_1, S_2} - V_{S_1, C_2} \nonumber \\
\label{eq:shell_shell_interaction}
              & = V_{B_1, B_2} - V_{C_1, C_2}   \nonumber \\
              &  - (V_{C_1, B_2} - V_{C_1, C_2}) - (V_{B_1, C_2} - V_{C_1, C_2}) \\ 
              & = V_{B_1, B_2} + V_{C_1, C_2} -  V_{C_1, B_2} - V_{B_1, C_2}\ . \nonumber
\end{align}
For two identical thick shells, the last two terms are identical. For
two identical shells of radius $R$ and thickness $h$, we can write 
\eq{eq:shell_shell_interaction} as
\begin{equation}
  \label{eq:identical_shell_shell_interaction}
  V_F(r) = V_{R R}(r) + V_{R-h, R-h}(r) -  2 V_{R, R-h}(r)\ .
\end{equation}
In order to proceed, we must integrate the Lennard-Jones potential
from \eq{eq:LJ12-6} over two balls of unequal radii. The two
integrations can be done analytically, but lead to rather lengthy
formulae which are given in Appendix~\ref{appendix}.
\eq{eq:ball_ball_interaction6} and \eq{eq:ball_ball_interaction12}
together then yield
\begin{equation}
  \label{eq:ball_ball_interaction}
  V_{R_1 R_2}(r) =  4 \epsilon \chi^2 \left( \sigma^6 V^{12}_{R_1 R_2}(r) - V^6_{R_1 R_2}(r) \right) 
\end{equation}         
which can be used in \eq{eq:identical_shell_shell_interaction} to
obtain the full potential $V_F(r)$ in analytical form. It diverges at
$r = 2 R$, has a minimum at intermediate distances and decays at large
distances as
\begin{equation}
  \label{eq:full_large_s}
  V_F(r) = - \pi^2 \chi^2 \epsilon \frac{16}{9} \nu^2 
(4 \nu^2 - 6 \nu \eta + 3 \eta^2) \frac{1}{(\eta s)^6} 
\end{equation}
which is the vdW interaction between two bodies of volume $4 \pi [R^3
- (R-h)^3]/3$ each. In the two-dimensional limit $h \to 0$, the full
potential $V_F$ between two hollow nanoparticles becomes the Girifalco
potential $V_G$ from \eq{eq:Girifalco} between two spherical shells
(where the dimensionless volume density $\chi$ relates to the
dimensionless area density $\tau$ by $\chi = \tau \sigma / h$).  In
general, the full potential $V_F$ is very complicated; in order to
gain some physical insight, it is useful to consider its Derjaguin
approximation.

\subsection{Derjaguin approximations}
\label{subsec:InteractionPotentialsC}

The Derjaguin approximation relates the interaction between curved
surfaces to the one between flat surfaces if the distance $r - 2 R$
between the curved surfaces is smaller than the radii of curvature
$R$, i.e.\ if $s \ll 2$.  The interaction energy per unit area between
two planar films a distance $z$ apart and each of thickness $h$ can be
easily calculated to be
\begin{align} 
\label{eq:flat_two_films}
W(z) & = \pi \chi^2 \epsilon \Bigg[
\frac{\sigma^6}{90} \left( \frac{1}{z^8} + \frac{1}{(z+2 h)^8} - \frac{2}{(z+h)^8} \right) \nonumber \\
& - \frac{1}{3} \left( \frac{1}{z^2} + \frac{1}{(z+2 h)^2} - \frac{2}{(z+h)^2} \right)
\Bigg]\ .
\end{align}
The Derjaguin approximation integrates over one of the two surfaces
and evaluates $W(z)$ for $z$ being the distance to the nearest point
on the other surfaces; for two sphere of equal radii $R$ it reads
\begin{equation}
  \label{eq:Derjaguin}
  V_D(r) = \pi R \int_{r - 2R}^{\infty} dz\ W(z)\ .
\end{equation}
Using \eq{eq:flat_two_films} in \eq{eq:Derjaguin} we find
\begin{align}
V_D(r) & = \pi^2 \chi^2 \epsilon \Big[
\frac{1}{1260 \eta^6} \Bigg( \frac{1}{(s-1)^7} + \frac{1}{(s-1+2 \nu / \eta)^7} \nonumber \\
\label{eq:derjaguin_two_films}
& - \frac{2}{(s-1+\nu/\eta)^7} \Bigg) \\
& - \frac{1}{6} \left( \frac{1}{s-1} + \frac{1}{s-1+2 \nu / \eta} 
- \frac{2}{s-1+\nu/\eta} \right) \nonumber
\Big]\ .
\end{align}

\begin{figure}
  \begin{center}
    \epsfig{file=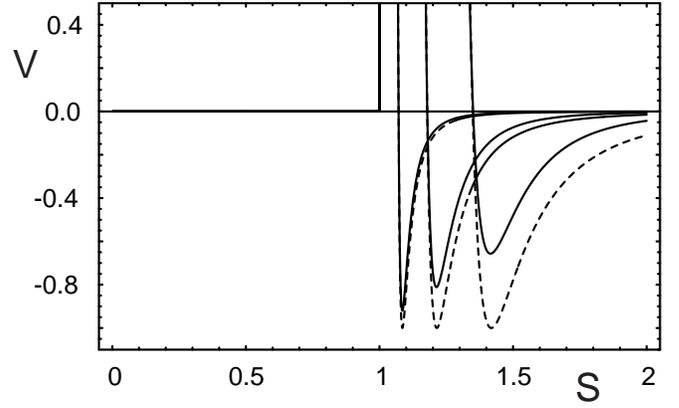,width=\columnwidth}
  \end{center}
  \caption{Interaction potential for single-walled nanoparticles:
    Girifalco potential (full) and Derjaguin approximation (dashed)
    for different values of $\eta = 2 R/\sigma$: from right to left
    $\eta = 2.05$ (buckyballs), $4$ and $10$. All potentials are
    scaled such that the Derjaguin approximation has its minimum at
    $-1$. With increasing $\eta$, the potential well of the full
    potential falls into the range described by the Derjaguin
    approximation.}
  \label{fig:ScaledPotentials}
\end{figure}

We now discuss two limits of this result. For $\nu \ll 1$, that is $h
\to 0$, we obtain the Derjaguin approximation for the Girifalco
potential from \eq{eq:Girifalco}:
\begin{equation}
  \label{eq:derjaguin_thin_films}
V_D^G(r) = \pi^2 \tau^2 \epsilon \left[ 
\frac{2}{45 \eta^8} \frac{1}{(s-1)^9} - \frac{1}{3 \eta^2} \frac{1}{(s-1)^3} \right]
\end{equation}
where we have used $\chi = \tau \sigma / h$ in order to get the
two-dimensional limit.  The same result follows formally by expanding
each of the two terms of the full potential from \eq{eq:Girifalco}
\emph{separately} around $s = 1$. The potential minimum in the
Derjaguin approximation is found by minimizing in
\eq{eq:derjaguin_thin_films} for $s$:
\begin{equation}
  \label{eq:min_thin_films}
  s_{0} = 1 + \left( \frac{2}{5} \right)^{\frac{1}{6}} \frac{1}{\eta},\ 
  V_{0} = - \frac{\sqrt{10}}{9} \pi^2 \tau^2 \epsilon \eta\ .
\end{equation}
The Derjaguin approximation is valid for $s \ll 2$; thus it will
describe the potential well correctly if $s_0 \ll 2$, that is if $\eta
\gg 1$. From \eq{eq:min_thin_films} we see that
in this limit of large $R$, the equilibrium distance of closest
approach, $2 R (s_0 - 1) = (2/5)^{1/6} \sigma = 2.98\ \AA$, is
independent of $R$.  Moreover it follows that in this limit, the
potential depth scales linearly in $R$.  In \fig{fig:ScaledPotentials}
we plot the Girifalco potential and its Derjaguin approximation
rescaled by $V_0$ and for $\eta = 2.05, 4$ and $10$. While in the
first case (buckyballs) the difference between full potential and
approximation is still considerable, for the last case it is already
very good.  Comparing values obtained numerically from the full
potential with the values from its Derjaguin approximation from
\eq{eq:min_thin_films}, we find that the approximation for $s_0$ is
very good even for buckyballs with $R = 3.55\ \AA$ ($\eta = 2.05$).
The approximation for $V_0$ deviates by $34\ \%$ for buckyballs, but
only by $10\ \%$ for $R = 13.74\ \AA$ ($\eta = 7.92$) and by $1\ \%$
for $R = 147.44\ \AA$ ($\eta = 84.98$).  

We now consider \eq{eq:Derjaguin} in the limit $\nu \gg 1$, that is $h
\gg \sigma$.  This is the Derjaguin approximation for a nanoparticle
which is completely filled:
\begin{equation}
 \label{eq:derjaguin_thick_films}
V^F_D(r) = \pi^2 \chi^2 \epsilon \left[ 
\frac{1}{1260 \eta^6} \frac{1}{(s-1)^7} - \frac{1}{6 (s-1)} \right]
\end{equation}
Equilibrium distance $s_0$ and potential depth $V_0$ again follow from
minimizing for $s$:
\begin{equation}
  \label{eq:min_thick_films}
  s_{0} = 1 + \left( \frac{1}{30} \right)^{\frac{1}{6}} \frac{1}{\eta},\ 
  V_{0} = - \frac{30^{1/6}}{7} \pi^2 \chi^2 \epsilon \eta\ .
\end{equation}
If we compare this result with \eq{eq:min_thin_films} for
single-walled nanoparticles, we see that apart from the numerical
prefactors and the change from area to volume density, for single- and
multi-walled nanoparticles the Derjaguin approximation basically
yields the same results for equilibrium distance $s_0$ and potential
depth $V_0$.  In particularly, in both cases the equilibrium distance
is independent of $R$ and on an atomic scale, and the potential depth
scales linearly in $R$. The reason for this agreement is that in the
situation of close approach described by the Derjaguin approximation,
the size of the gap between the particles is approximately $\sigma$.
Since the vdW interaction between two films saturates on the same
length scale, it doesn't really matter how many walls there are.
Moreover there is a cancellation of two effects: in the case of
multi-walled particles, more matter is present and the vdW energy
increases. However, in our description of thick walls, matter is now
distributed uniformely in space and the effect of the first layer is
smeared out, so the vdW energy decreases. In reality both descriptions
are idealizations; however, since they essentially yield the same
result for $s_0$ and $V_0$, the overall description of the potential
well should be sufficient. Note that in both cases, the temperature
scale $V_0 / k$ (where $k$ is the Boltzmann constant) is approximately
$8 \eta T_R$, where $T_R$ denotes room temperature. Thus if a
gas-liquid coexistence exists for the systems under consideration, it
will do so at several thousand Kelvin. In fact the detailed treatments
for buckyballs predict the gas-liquid coexistence to exist close to
2000 K.

\section{Material properties}
\label{sec:MaterialProperties}

A solid forms at high densities or low temperatures since then
close-packed particles can benefit from the attractive interaction
without losing too much entropy. Many of the physical properties of
solids are essentially determined by the properties of the potential
well. We showed above that for large radius, $R$, the potential well
is well described by the Derjaguin approximation. We now proceed to
predict the material properties of such solids. By using the Derjaguin
approximations rather than the full potential, we are able to derive
analytical formulae which show the scaling with the different
parameters involved; although they are strictly valid only for $\eta =
2 R / \sigma \gg 1$, our results predict the right trends and orders
of magnitude even for small radius. For more precise results for small
$R$ (relevant to buckyballs), it is easy to treat the full problem
numerically \cite{n:giri92}.  Because the Derjaguin approximation
resulted in very similar results for the potential well of single- and
multi-walled nanoparticles, it is sufficient for the following to use
the potential for the former case, i.e.\ \eq{eq:min_thin_films}.

Fullerite has several unusual material properties.  It has a very high
heat of sublimation ($40.1$ kcal/mol, 5 times larger than typical for
vdW solids) \cite{n:pan91} and it is the softest carbon structure; its
volume compressibility $7 \times 10^{-12}$ cm$^2$/dyn is 3 and 40
times the values for graphite and diamond, respectively
\cite{n:fisc91}. We now investigate how these quantities scale for
different $R$ and $h$.  The calculations proceeds as for Lennard-Jones
solids \cite{a:ashc76}.  We assume the crystal to be face centered
cubic (fcc). The energy per particle is
\begin{align}
  \label{eq:energy_per_particle}
  u & = \frac{1}{2} \sum_{i} n_i V(r_i) = \frac{1}{2} \Big(
      12\ V(r) \nonumber \\
& + 6\ V(\sqrt{2}\ r) + 24\ V(\sqrt{3}\ r) + \dots \Big)
\end{align}
where the sum is over all fcc Bravais lattice sites (except the origin
which is occupied by the particle itself) and we grouped the lattice
sites into shells $\{ i \}$ of equal distance $\{ r_i \}$ to the
origin, where shell $i$ contains $n_i$ different sites. We explicitly
give the nearest, next-nearest and next-next-nearest neighbor shells.
Here $r$ is measured in units of $r_{nn}$, the nearest neighbor
distance in the crystal, which follows from $du(r_{nn}) / dr = 0$. In
the following we will use the nearest neighbor approximation, $u
\approx 6 V(r)$.  Then $r_{nn} = 2 R s_0$ and $u = 6 V_0$ with $s_0$
and $V_0$ from \eq{eq:min_thin_films}. The lattice constant is then
given by $a = \sqrt{2} r_{nn}$.

The heat of sublimation per mol is simply $\Delta H = N_A u(r_{nn}) =
6 N_A V_0$ with Avogadro's number $N_A = 6.02 \times 10^{23}$
mol$^{-1}$. Therefore
\begin{equation}
\Delta H = \frac{2 \sqrt{10}}{3} \pi^2 N_A \tau^2 \epsilon \eta\ .
\end{equation}
Since in the nearest neighbor approximation the heat of sublimation
depends only on the direct interaction between the particles and not
on any geometrical aspect of the crystal lattice, it scales linearly
with $R$. Thus we find that the heat of sublimation, which is already
unusually high for buckyballs, increases even more with radius $R$.
Using $\tau = 4.56$ and $\eta = 2.05$ for buckyballs yields $\Delta H
= 58.46$ kcal/mol; given that we use the nearest neighbor and the
Derjaguin approximation (which is valid for $\eta \gg 1$), the
agreement with the experimental value of $40.1$ kcal/mol
\cite{n:pan91} is surprisingly good.

The bulk modulus follows as \cite{a:ashc76}
\begin{align}
  \label{eq:bulk_modulus}
  B & = -V \frac{\partial P}{\partial V}
    = \frac{\sqrt{2}}{9 r_{nn}} \frac{\partial^2 u}{\partial r^2}(r_{nn}) \\
    & = \frac{\sqrt{2} 6}{9 (2 R)^3 s_0} \frac{\partial^2 V_D}{\partial s^2}(s_0)
    = \frac{20^{5/6} \pi^2 \tau^2}{s_0} \frac{\epsilon}{\sigma^3}\ . \nonumber
\end{align}
For $\eta \gg 1$, $s_0 \approx 1$ and B becomes independent of $R$.
Therefore its scale must be set by $\epsilon/\sigma^3$ from
dimensional considerations.  The volume compressibility follows as
$\beta = 1 / B$.  Extrapolating this to buckyballs, we find $\beta =
3.65 \times 10^{-12}$ cm$^2$/dyn, which again is in surprisingly good
agreement with the experimental value $\beta = 7 \times 10^{-12}$
cm$^2$/dyn \cite{n:fisc91}.

The energy per area needed to cleave the crystal into two along
lattice planes perpendicular to a given direction is twice the surface
energy $\gamma$ for this direction. For simplicity we will consider
only the (111)-direction. Before doing so, we first discuss $\gamma$
for the layered material. Using the nearest neighbor approximation, we
have to determine the energy per area at the equilibrium separation of
two planar sheets. Taking the limit $h \to 0$ in
\eq{eq:flat_two_films}, using $\chi = \tau/h$ and minimizing for
distance $z$ yields
\begin{equation}
  \label{eq:gamma_graphite}
  \gamma_G = \frac{3}{5} \pi \tau^2 \frac{\epsilon}{\sigma^2}\ .
\end{equation}
Again the combination $\epsilon / \sigma^2$ follows from dimensional
considerations.  For carbon we find $\gamma = 150$ erg/cm$^2$, which
compares quite well with the experimental value of $\gamma = 130$
erg/cm$^2$ for graphite \cite{n:giri56}.

We now consider the (111)-direction of the fcc-solid.  Since a given
molecule in a corresponding lattice plane has three nearest neighbors
in the adjacent plane, $\gamma$ can be estimated to be $3 V_0/2$ times
the area density in a close-packed plane with nearest neighbor
distance $r_{nn}$, $2 / \sqrt{3} r_{nn}^2$:
\begin{equation}
\gamma = \frac{\sqrt{10}}{3 \sqrt{3}} 
\pi^2 \tau^2 \frac{1}{s_0^2 \eta} \frac{\epsilon}{\sigma^2}
= \frac{5 \sqrt{10}}{9 \sqrt{3}} \frac{1}{s_0^2 \eta} \gamma_G\ .
\end{equation}
For $\eta \gg 1$, $s_0 \approx 1$ and the surface energy $\gamma$
scales inversely with $R$.  Extrapolating to buckyballs, we find
$\gamma = 115$ erg/cm$^2$, that is a value smaller than the surface
energy $\gamma_G$ of graphite. Although the interaction becomes
stronger as $R$, the number of vdW contacts decreases as $1/R^2$ and
the geometrical effects dominates. With increasing $R$, it hence
becomes easier to cleave the crystal. Note however that this does not
lead to an enhanced roughening of the crystal surface. In fact the
roughening temperature scales as $\gamma a^2 \sim R$ with $a \sim R$
being the lattice constant \cite{a:safr94}. Since the melting
temperature scales with the potential depth which in turn scales with
$R$, both temperatures scale with $R$ and the roughening temperature
remains a certain fraction of the melting temperature with varying
$R$, as it usually is.

\section{Phase behavior}
\label{sec:PhaseBehavior}

For the simple Lennard-Jones interaction of \eq{eq:LJ12-6},
we can rescale the potential with the well depth, $\epsilon$,
and the distance $r$ with the hard core diameter $\sigma$,
to write
\begin{equation}
  \label{eq:LJ12-6_rescaled}
  V_{LJ}(r) = 4 \left[ \left(\frac{1}{r}\right)^{12} 
                  - \left(\frac{1}{r}\right)^6 \right]
\end{equation}
which has no other free parameters. The resulting phase diagram in
reduced temperature $t = kT / \epsilon$ and volume density $\Phi = \pi
\rho \sigma^3 / 6$ (where $\rho$ is density and $\sigma$ hard core
diameter) is universal; the phase diagrams for all Lennard-Jones
systems are identical when plotted in units of $t$ and $\Phi$
(\emph{law of corresponding states}). In
\sec{subsec:InteractionPotentialsB} we derived effective interaction
potentials for single- and multi-walled nanoparticles. Single-walled
nanoparticles interact via the Girifalco potential $V_G$ from
\eq{eq:Girifalco}.  Although derived from the Lennard-Jones potential
from \eq{eq:LJ12-6} for two single atoms, it involves an additional
length scale, the particle diameter $2 R = \eta \sigma$. Multi-walled
nanoparticles interact with the complicated potential $V_F$ given by
Eqs.~(\ref{eq:identical_shell_shell_interaction},\ref{eq:ball_ball_interaction}).
Here another length scale enters, the wall thickness $h = \nu \sigma$.
Since they involve more than one length scale, the potentials $V_G$
and $V_F$ do not lead to a law of corresponding states and their phase
behavior needs closer inspection.

In \sec{subsec:InteractionPotentialsC} we showed that for large $R$,
the Derjaguin approximations given in \eq{eq:derjaguin_thin_films} and
\eq{eq:derjaguin_thick_films} for $V_G$ and $V_F$, respectively, give
good approximations of the potential well.  In
\sec{sec:MaterialProperties} this model was used to predict certain
physical properties of solids which are essentially determined by the
properties of the potential near its minimum. In regard to phase
behavior, the situation is more complicated, since now the entire
range of the potential matters. However, recent work
\cite{n:bolh94,n:teje,a:daan}
suggests that for an isotropic interaction potential with hard core
repulsion and some attractive component for which potential depth and
hard core diameter have been used to scale temperature and density to
reduced units, the topology of the phase diagram will be essentially
determined by \emph{one} additional feature of the effective
potential, namely its \emph{interaction range}. This quantity is well
defined only for square well potentials where it is the ratio of well
width to hard core diameter.  For continuous potentials, the following
definition has been employed \cite{n:teje}: $\delta =
(r_1-r_0)/r_0$ where $r_0$ is the position of the potential minimum
and $r_1$ the distance at which the potential has fallen to 1/100 of
the potential depth at $r_0$. Note that the interaction range $\delta$
is independent of the potential depth $\epsilon$.  For the
Lennard-Jones potential given in \eq{eq:LJ12-6_rescaled}, $\delta =
1.42$, and for the Girifalco potential for buckyballs given in
\eq{eq:Girifalco}, $\delta = 0.83$.  In the framework of a variational
scheme for the Double-Yukawa potential, it was found that the
gas-liquid coexistence disappears for $\delta \lesssim 0.4$, and that
an isostructural solid-solid coexistence appears for $\delta \lesssim
0.02$ \cite{n:teje}. Similar values where found for the
square well potential both in a simple van der Waals theory
\cite{a:daan} and in extensive MC simulations
\cite{n:bolh94}.

\begin{figure}
  \begin{center}
    \epsfig{file=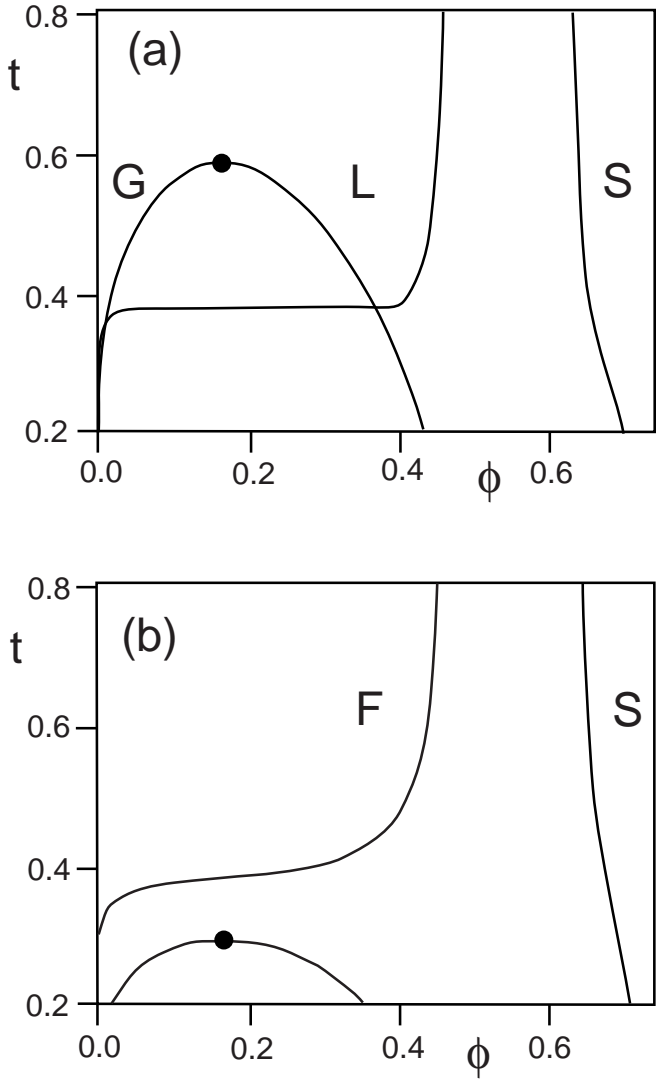,width=\columnwidth}
  \end{center}
  \caption{Phase diagrams (a) with and (b) without gas-liquid coexistence
    as a function of reduced temperature $t = k T / \epsilon$ and
    volume fraction $\phi = \pi \rho \sigma^3 / 6$. G, L, S and F
    denote gaseous, liquid, solid and fluid phases, respectively.  In
    (b) the attractive interaction potential is so short-ranged that the
    gas-liquid coexistence (although existing by itself) is suppressed
    in the overall phase diagram by the fluid-solid coexistence.  For
    a Double-Yukawa potential, the crossover between the two
    topologies of the phase diagram occurs when the interaction range
    decreases below the critical value $\delta \approx 0.4$.}
  \label{fig:PhaseDiagrams}
\end{figure}

The physical reason for the disappearance of the gas-liquid
coexistence is well known. The fluid-solid phase transition is driven
mainly by changes in free volume: at high densities, the solid becomes
more favorable since its ordered structure provides more space for
fluctuations (therefore hard sphere systems exhibit a entropy-driven
fluid-solid transition). In contrast, the gas-liquid coexistence is
driven by the attractive interaction: at low temperatures, the fluid
phase separates so that the gas and liquid phases can profit
entropically and energetically, respectively.  The critical
temperature for the fluid-fluid coexistence scales linearly with the
depth of the potential and also depends on its higher moments. Even if
the potential is only slightly attractive, the critical point will
exist by itself.  However, the corresponding gas-liquid coexistence
will survive in the overall phase diagram only if the critical
temperature lies above the temperature at which the fluid coexists
with the solid at the critical density.  In \fig{fig:PhaseDiagrams} we
schematically depict the two possible outcomes which we have to
consider in this work.

The calculation of phase diagrams from various interaction potentials
has reached a high level of sophistication and different schemes are
readily available for a detailed analysis (compare \cite{n:hase99} for
an overview of different methods applied in the case of buckyballs).
In the following we ask how does the gas-liquid coexistence disappear
from the phase diagram of hollow nanoparticles as a function of $R$
and $h$.  For this purpose, it is sufficient to adopt the simple
criterion which was derived within the framework of a sophisticated
variational scheme for the Double-Yukawa potential: the gas-liquid
coexistence disappears if the interaction range $\delta$ falls below
$\delta \approx 0.4$. The basic justification for our approach lies in
the fact that the interaction potentials under consideration can be
fit well by the Double-Yukawa potential. This potential has in fact
three parameters \cite{n:teje}, one of which scales the potential
depth.  The remaining two can be used to adjust the position of the
potential minimum and the interaction range. The zero crossing of the
potential can be identified with the hard core diameter and for the
Double-Yukawa potential equals one by definition. For colloids as well
as for the hollow nanoparticles discussed in this work, the position
of the potential minimum should always be close to the hard core
diameter, that is little larger than one. Therefore we are left with
only one degree of freedom, which in our approach can be used to fit
the interaction range $\delta$.  The interaction range for the full
potentials $V_G$ and $V_F$ can be determined numerically. The
corresponding Derjaguin approximations can be used to gain some
physical insight into the numerical results. In fact, for very large
$R$ the Derjaguin approximation should correctly describe not only the
part of the potential which contains the potential minimum, but also
the part which contains the interaction range.

\begin{figure}
  \begin{center}
    \epsfig{file=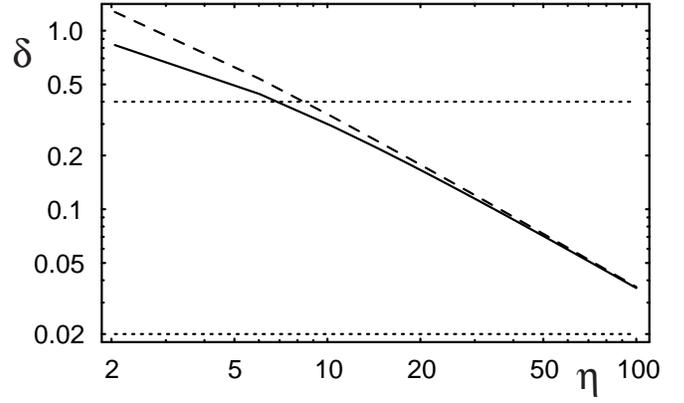,width=\columnwidth}
  \end{center}
  \caption{Double logarithmic plot of the interaction range $\delta$
    of single-walled nanoparticles as a function of $\eta = 2 R /
    \sigma$ for the Girifalco potential (full) and its Derjaguin
    approximation (dashed). The dotted lines at $\delta = 0.4$ and
    $\delta = 0.02$ separate the long-, intermediate- and short-ranged
    regimes.  Buckyballs with $\eta = 2.05$ still have the gas-liquid
    coexistence which disappears only for $\delta < 0.4$, that is
    $\eta > 7$ ($R > 12\ \AA$).}
  \label{fig:InteractionRangeGirifalco}
\end{figure}

We first discuss single-walled nanoparticles. The Derjaguin approximation
to the Girifalco potential $V_G$ is given in \eq{eq:derjaguin_thin_films}.
Combining it with the result for its potential depth, \eq{eq:min_thin_films},
gives
\begin{equation}
  \label{eq:derjaguin_thin_films_rescaled}
V_D^G(r) = V_0 \left[ 
\frac{2}{5 \sqrt{10}} \frac{1}{(\eta (s-1))^9} 
- \frac{3}{\sqrt{10}} \frac{1}{(\eta (s-1))^3} \right]\ .
\end{equation}
Thus we find that for large $R$, the particle diameter, $2 R = \eta
\sigma$, simply serves to rescale the separation distance. With $s_0$
from \eq{eq:min_thin_films} and $s_1 = 1 + 4.56 / \eta$ determined
numerically, we find
\begin{equation}
  \label{eq:interaction_range}
  \delta = \frac{3.70}{\eta + (2/5)^{1/6}} 
         = \frac{3.70}{\eta} + {\cal O}(\frac{1}{\eta^2})\ .
\end{equation}
Thus we conclude that the dimensionless interaction range $\delta$
scales inversely with $R$ for large $R$. This result can be understood
as follows: on an absolute scale, the interaction range is essentially
constant since it corresponds to the range of the Lennard-Jones
interaction between atoms. It is made dimensionless by rescaling with
the position of the effective potential minimum, which is fixed by the
Lennard-Jones potential close to $2 R$; therefore the dimensionless
interaction range $\delta$ scales inversely with $R$.  In
\fig{fig:InteractionRangeGirifalco} we plot the interaction range of
the Girifalco potential and the result from the Derjaguin
approximation, \eq{eq:interaction_range}.  Although the curves
collapse onto each other for large $\eta$, for $\eta = 2.05$ they
disagree considerably ($\delta = 0.83$ versus $\delta = 1.27$).
Guerin \cite{n:guer98b} fit the Girifalco potential to a Double-Yukawa
potential for which he found $\delta = 0.56$.  The difference between
this value for the fit and $\delta = 0.83$ for the Girifalco potential
arises since the fitting procedure does not preserve the value of the
interaction range. The approach used in this work amounts to using the
three free parameters of the Double-Yukawa potential to fit it to the
potential depth $\epsilon$, the position of the minimum $r_0$ and the
interaction range $\delta$ of the potential under investigation, thus
preserving the value of the interaction range.
\fig{fig:InteractionRangeGirifalco} shows that for $\eta \approx 7$,
the full potential has $\delta \approx 0.4$ and the liquid-gas coexistence
is expected to disappear. This corresponds to $R \approx 12\ \AA$,
i.e.\ $C_{690}$. On the basis of these results, we have to conclude
that buckyballs $C_{60}$ are in fact rather far away from loosing
their gas-liquid coexistence.

We now proceed to discuss multi-walled nanoparticles. The Derjaguin
approximation is given in \eq{eq:derjaguin_two_films}. Since it
involves one more length scale, it cannot be treated in the same way
as the Derjaguin approximation for the Girifalco potential. However,
we showed above that the potential saturates for large $h$. The limit
$h \gg \sigma$ is given in \eq{eq:derjaguin_thick_films}.  Combining this
with the result for the potential depth, \eq{eq:min_thick_films},
gives
\begin{equation}
 \label{eq:derjaguin_thick_films_rescaled}
V^F_D(r) = V_0 \left[ 
\frac{7}{1260\ 30^{1/6}} \frac{1}{(\eta (s-1))^7} 
- \frac{7}{6\ 30^{1/6} (\eta (s-1))} \right]\ .
\end{equation}
As in the case for single-walled nanoparticles, $2 R = \eta \sigma$
rescales the particles separation. With 
$s_0$ from \eq{eq:min_thick_films} and $s_1 = 1 + 66.18 / \eta$
determined numerically, we now find
\begin{equation}
  \label{eq:interaction_range_thick}
  \delta = \frac{65.61}{\eta + (1/30)^{1/6}} 
         = \frac{65.61}{\eta} + {\cal O}(\frac{1}{\eta^2})\ .
\end{equation}
Thus the dimensionless interaction range $\delta$ scales inversely
with $R$ in both cases, although it has a larger prefactor in the
multi-walled case, so that the gas-liquid coexistence disappears at
larger values of $R$ for larger wall thickness $h$.

\begin{figure}
  \begin{center}
    \epsfig{file=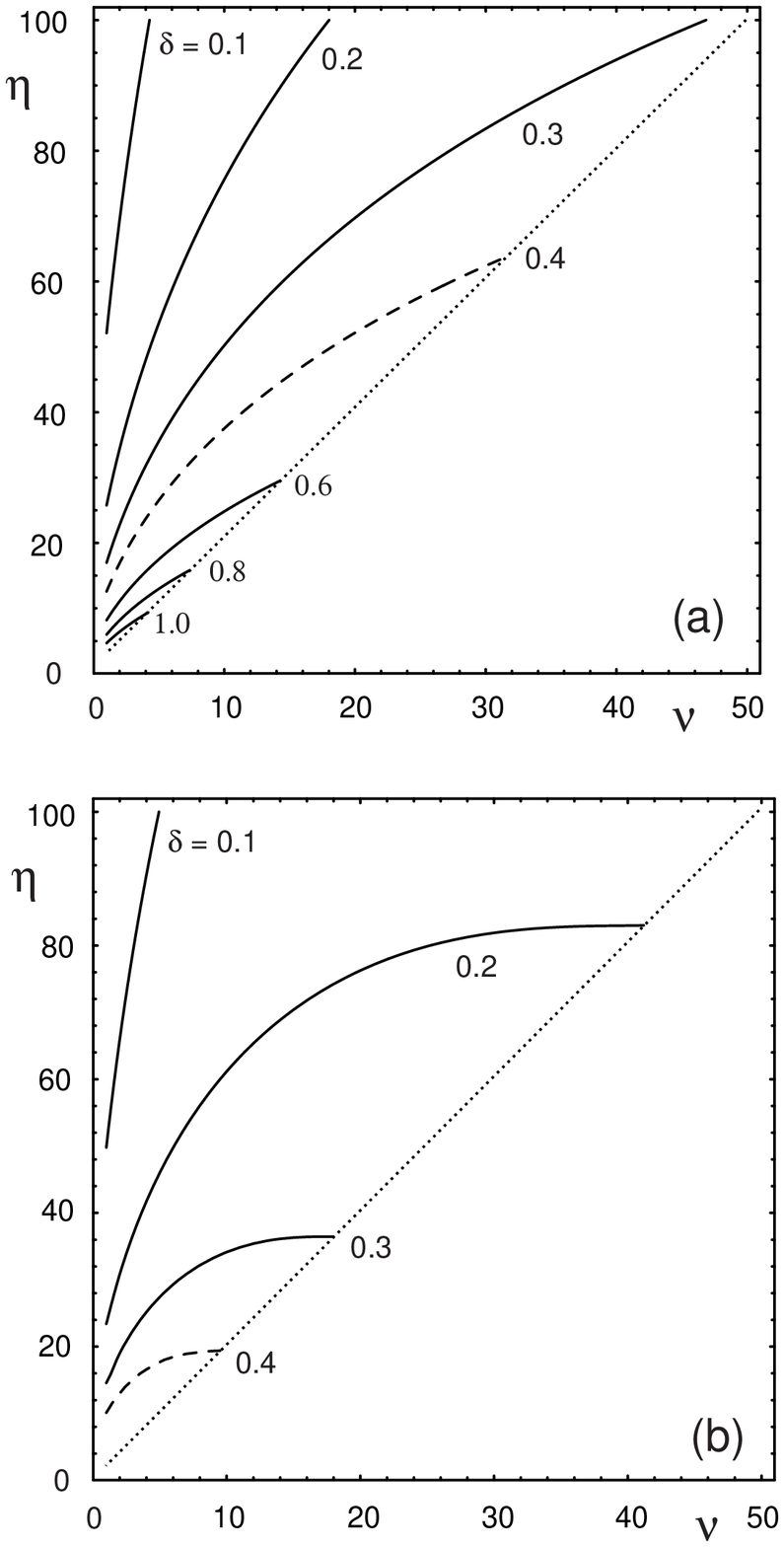,width=\columnwidth}
  \end{center}
  \caption{Interaction range $\delta$ of
    multi-walled nanoparticles as a function of $\eta = 2 R / \sigma$
    and $\nu = h / \sigma$ for (a) Derjaguin approximation and (b)
    full potential. The dotted line delimits the allowed region $h \le
    R$, that is $\eta \ge 2 \nu$. The dashed line marks $\delta =
    0.4$. Only the region between the dotted and the dashed lines has
    a gas-liquid coexistence, whose size is overestimated in the
    Derjaguin approximation. Filled nanoparticles ($\eta = 2 \nu$) do
    not have a liquid phase for $\eta > 20$ ($R > 35\ \AA$).}
  \label{fig:InteractionRange}
\end{figure}

In \fig{fig:InteractionRange} we plot the numerical results for the
interaction range $\delta$ of the full potentials and their Derjaguin
approximations. Different values of the interaction range $\delta$ are
shown as isolines in the $(\nu, \eta)$-plane, and the critical
interaction range $\delta_c = 0.4$ is drawn as a dashed line.  The
gas-liquid coexistence exists only in the region below the dashed
line. Both diagrams agree with the general trends which were explained
above: the coexistence disappears for large $R$ and small $h$, and the
critical line levels off for large $h$.  However, the Derjaguin
approximation overestimates the interaction range, the real region
with coexistence being in fact much smaller.  From the results for the
full potentials, we can conclude that the coexistence disappears for
$\eta \approx 7$ ($R \approx 12\ \AA$) for single-walled and for $\eta
\approx 20$ ($R \approx 35\ \AA$) for filled nanoparticles,
respectively; the crossover with increasing wall thickness is rapid
and on an atomic length scale. In particular we now can predict that
buckyballs (single-walled, $R = 3.55\ \AA$, $\eta = 2.05$, $h = \nu =
0$) will have a gas-liquid coexistence, while typical metal
dichalcogenides nanoparticles (multi-walled, $R = 600\ \AA$, $\eta =
346$, $h = 93\ \AA$, $\nu = 27$) will almost certainly have none.

\section{Elastic deformations}
\label{sec:Elasticity}

We now briefly discuss at which values of $R$ and $h$ the effects
investigated above may be modified by the deformability of the particles.
Although fullerene-like material interacts only through vdW forces,
the resulting elastic deformations can be substantial, as has been
shown both experimentally and theoretically for carbon nanotubes
\cite{n:ruof93}. A detailed treatment of the elastic
deformation of hollow nanoparticles is difficult and relatively
unexplored. Here we adopt a simple scaling approach which has been
used before to investigate the mechanical stability of hollow
nanoparticles in tribological applications \cite{uss:schw00d}.

\begin{figure}
  \begin{center}
    \epsfig{file=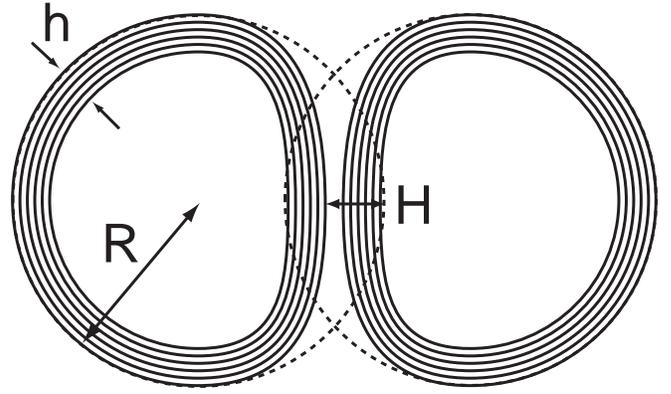,width=\columnwidth}
  \end{center}
  \caption{Schematic drawing of two multi-walled hollow nanoparticles 
    deformed due to vdW adhesion. $H$ is the indentation. It follows
    geometrically that the contact area scales with $R H$.}
  \label{fig:Deformation}
\end{figure}

We ask when will considerable elastic deformations of hollow
nanoparticles occur due to vdW interactions as a function of $R$ and
$h$. We consider two hollow nanoparticles adhering to each other due
to their mutual vdW interaction as depicted in \fig{fig:Deformation}.
The hollow nanoparticle is assumed to be an elastic shell with
prefered radius $R$. Its deformation energy has two parts: bending
energy characterized by the bending constant $\kappa$ and stretching
energy characterized by the two-dimensional Young modulus $G$. If an
external force is applied, the balance between these two contributions
leads to a localization of the deformation \cite{b:land70}.  For small
adhesive load, the shell flattens in a contact region and the overall
deformation energy scales as
\begin{equation}
  \label{eq:loss}
  E_{def} \sim \frac{G^{1/2} \kappa^{1/2}}{R} H^2
\end{equation}
where $H$ is the indentation (compare \fig{fig:Deformation}). The vdW
energy gained on deformation can be estimated as follows: apart from
some numerical prefactor, the adhesion energy per area is essentially
$\chi^2 \epsilon / \sigma^2$ (this follows from minimizing for
distance $z$ in \eq{eq:flat_two_films}; the limit of single walls has
been given in \eq{eq:gamma_graphite}). For small $H$, it follows from
simple geometry that the area over which the two shells approach each
other scales with $R H$ (compare \fig{fig:Deformation}). Therefore we
have
\begin{equation}
  \label{eq:gain}
  E_{vdW} \sim \frac{\chi^2 R \epsilon}{\sigma^2} H\ .
\end{equation}
For single-walled nanoparticles, $\chi$ has to be replaced by $\tau$,
which however has nearly the same value.  We now can solve $E_{def} =
E_{vdW}$ for indentation $H$. We then ask for the critical radius
$R_c$ at which the indentation becomes considerable, i.e.\ of the
order of the smallest length scale in our problem, which is $\sigma$.
This yields
\begin{equation}
  \label{eq:indentation}
  \eta_c = \frac{2 R_c}{\sigma} \sim \left( 
\frac{G^{1/2} \kappa^{1/2} \sigma}{\chi^2 \epsilon} \right)^{1/2}\ .
\end{equation}
For single-walled fullerenes, the value for the bending rigidity
$\kappa$ can be extracted from molecular calculations to be $\kappa =
1.6 \times 10^{-12}$ erg.  The value for the two-dimensional Young
modulus $G$ can be extracted from the elastic moduli of graphite as $G
= 3.6 \times 10^5$ erg/cm$^2$. Then we find $\eta_c = 16.5$, that is
considerable deformation is expected for $R > 30\ \AA$.  For
single-walled nanoparticles made from metal dichalcogenides MoS$_2$
and WS$_2$, $G$ is smaller by a factor $4$ and $\kappa$ is larger by a
factor $10$, thus $\eta_c$ increases slightly (by a factor $1.3$).

For nanoparticles with a few walls ($\sigma < h < R$), one can show
in the framework of linear elasticity theory that $\kappa = C_{11} h^3
/ 12$ and $G = C_{11} h$ where $C_{11}$ is the in-plane stretching
elastic constant of the corresponding layered material
\cite{n:srol94}. Using these relations yields
\begin{equation}
  \label{eq:multi_walled_indentation}
   \eta_c = \frac{2 R_c}{\sigma} \sim \left( \frac{C_{11} \sigma^3}{\chi^2 \epsilon} \right)^{1/2} 
                         \frac{h}{\sigma}\ .
\end{equation}
Thus the critical radius $R_c$ scales linearly with the wall thickness
$h$. The values for $C_{11}$ are $1060$, $238$ and $150 \times
10^{10}$ erg/cm$^3$ for C, MoS$_2$ and WS$_2$, respectively
\cite{b:land91}. For WS$_2$ particles with $h = 100\ \AA$, the
critical radius following from \eq{eq:multi_walled_indentation} is
well above $1000\ \AA$. 

We found above that the gas-liquid coexistence disappears for $R
\approx 12\ \AA$ for single-walled and for $R \approx 35\ \AA$ for
filled nanoparticles. Our estimates for the onset of elastic effects
is at least of the same order of magnitude.  Therefore we conclude
that the elastic effects will not affect hollow nanoparticles which do
have a fluid-fluid coexistence. In general, the effect of wall
thickness $h$ of multi-walled nanoparticles is much stronger on the
elastic response than it is on the phase behavior. Therefore
multi-walled variants are favorable if one wants to make use of the
unusual properites of hollow nanoparticles without being restricted by
elastic effects.

\section{Conclusion}
\label{sec:Conclusion}

In this work, we predicted the phase behavior and material properties
of hollow nanoparticles as a function of radius $R$ and wall thickness
$h$. The synthesis of hollow nanoparticles is a rapidly developing
field, driven by the special properties of particles in the nanometer
range which may allow many new applications to be developed in the
future \cite{n:review}.  Although our analysis is aimed at hollow
nanoparticles made from anisotropic layered material like carbon
(fullerenes) or metal dichalcogenides (inorganic fullerenes), it could
also be applied to colloidal or biological examples of hollow
nanoparticles.  Our starting point was the Lennard-Jones interaction
between single atoms, which we integrated over the appropriate
geometries in order to obtain effective potentials for the vdW
interaction between single- and multi-walled nanoparticles. The
subsequent analysis was twofold: Derjaguin approximations for the
effective potentials provided the correct scaling laws and physical
insight, while numerical investigations of the full potentials allowed us
to make quantitative predictions.

We first showed that crystals of hollow nanoparticles have
unusual material properties. Since the effective contact area and
therefore the potential depth scales linearly with radius $R$ for large
$R$, their heat of sublimation and surface energy will scale linearly
and inversely with $R$, respectively.  This means that with increasing
$R$, it will become more difficult to melt, but easier to cleave the
crystal. The thickness $h$ has little influence here since the vdW
interaction saturates over the atomic length scale of the
Lennard-Jones potential under close-packed conditions.  For reasonable
values of $R$, we predicted values for the heat of sublimation and surface
energy which for vdW solids are unusually high and low, respectively;
this was already found experimentally for fullerite, but should be
more pronounced for larger fullerenes and inorganic fullerenes.

Our discussion of phase behavior centered around the existence of a
liquid phase and the concept of the interaction range.  We argued that
after scaling out hard core diameter and potential depth, the
interaction range is the relevant feature of an effective potential
which determines phase behavior. We showed that for large radius $R$,
the interaction range scales inversely with $R$, and again is little
affected by thickness $h$. The numerical analysis supported this
finding and allowed to calculate the interaction range as a function
of $R$ and $h$. Using recent work on the Double-Yukawa potential,
which can be fit well to our potentials, we could identify the
disappearance of the liquid phase with a dimensionless interaction
range $\delta \approx 0.4$. We then found that the liquid phase will
disappear for $R \approx 12\ \AA$ for single-walled and for $R \approx
35\ \AA$ for filled nanoparticles. Although several theoretical
studies investigated the phase behavior of buckyballs as a likely
candidate for the non-existence of the liquid phase in a non-colloidal
system, we conclude that buckyballs with $R = 3.55\ \AA$ are in fact
rather far away from losing their liquid phase.

Finally, we showed that the wall thickness $h$ has a strong effect on the
elastic properties of hollow nanoparticles.  Using scaling arguments
for elastic shells with prefered radius $R$ and the known scaling of
the elastic moduli with $h$, we predicted that elastic effects should
not interfere with the phase behavior of those (inorganic) fullerenes which
do have a liquid phase.

Despite the widespread theoretical interest in the phase behavior of
buckyballs, little experimental work has been done on the phase
behavior of hollow nanoparticles. Since the potential depth scales
linearly with the radius $R$, the corresponding energy scales are
large and experiments have to be conducted at temperatures of several
thousand Kelvin, where their mechanical stability becomes a limiting
factor.  However, the temperature scale can in principle be lowered by
up to two orders of magnitude by dispersing the nanoparticles in a
fluid medium of high dielectric constant \cite{a:isra92}.  Regarding
hollow nanoparticles which are impermeable to solvent, it would not be
difficult to extend our analysis to the case of vdW interactions
between spatial regions with three instead of two different dielectric
constants. Our analysis should be directly applicable to the vdW
interaction in systems with hollow polyelectrolyte and silica shells,
whose walls are permeable to solvent \cite{n:dona98,n:caru98}.
However, in this case additional effects like electrostatic
interactions might have to be taken into account. Moreover, elastic
and even plastic effects (low yield strength due to non-covalent
bonding in the shell) are expected to be more prominent for these
systems.

\acknowledgments We wish to thank S.~Komura and R.~Tenne for useful
discussions. USS gratefully acknowledges support by the Minerva
Foundation. SAS thanks the Schmidt Minerva Center for Supramolecular
Architecture, the Israel Academy of Sciences and the Research Center
on Self Assembly of the Israel Science Foundation for support.

\begin{appendix}
\section{Integration of Lennard-Jones potential over two balls of unequal radii}
\label{appendix}

In order to integrate the Lennard-Jones potential over two balls of
unequal radii $R_1$ and $R_2$ and center-to-center distance $r$, it is
helpful to adopt the following coordinate system: consider a sphere of
radius $R$ centered at $z = r$ on the z-axis. This sphere can be
decomposed into caps of spherical shells of radius $r'$ centered
around the origin.  Integrating out the two angles on the spherical
caps leaves us with the following integral \cite{a:maha76}:
\begin{equation}
  \label{eq:IntegrationSpheres}
\int_{sphere} dV f(r') = \frac{\pi}{r} \int_{r-R}^{r+R} dr' {r'}^2 \frac{R^2 - (r-r')^2}{r'} f(r')
\end{equation}
E.g.\ for $f(r) = 1$ we obtain the sphere volume $4 \pi R^3 / 4$.  To
integrate a given $f(r)$ over two spheres, the same integral has to be
used twice, where in the second integration $f(r)$ is the result of the first
integration.  For $f(r) = 1/{r}^6$ we find \cite{a:maha76}
\begin{align}
  \label{eq:ball_ball_interaction6}
  V^6_{R_1 R_2}(r) & = \frac{\pi^2}{6} \Bigg( \frac{2 R_1 R_2}{r^2 - (R_1 + R_2)^2}
    + \frac{2 R_1 R_2}{r^2 - (R_1 - R_2)^2} \nonumber \\
    & + \log \frac{r^2 - (R_1 + R_2)^2}{r^2 - (R_1 - R_2)^2} \Bigg)
\end{align}   
and for $f(r) = 1/{r}^{12}$
\begin{align}
V^{12}_{R_1 R_2}(r) & = \frac{16 \pi^2 R_1^3 R_2^3}{4725 (r^2 - (R_1 - R_2)^2)^7 (r^2 + (R_1 + R_2)^2)^7} \nonumber \\
& \Big[ -35\,{\left( R_1^2 - R_2^2 \right) }^6\,\left( 5\,R_1^4 + 14\,R_1^2\,R_2^2 + 5\,R_2^4 \right)  \nonumber \\
& - 700\,{\left( R_1 - R_2 \right) }^4\,{\left( R_1 + R_2 \right) }^4\,\left( R_1^2 + R_2^2 \right) \, \nonumber \\
&   \left( R_1^4 + 10\,R_1^2\,R_2^2 + R_2^4 \right) \,r^2 \nonumber \\
& + 56\,{\left( R_1^2 - R_2^2 \right) }^2\,\big( 70\,R_1^8 - 49\,R_1^6\,R_2^2 - 762\,R_1^4\,R_2^4 - \nonumber \\
&     49\,R_1^2\,R_2^6 + 70\,R_2^8 \big) \,r^4 \nonumber \\
\label{eq:ball_ball_interaction12}
& - 4\,\left( R_1^2 + R_2^2 \right) \,\big( 875\,R_1^8 - 11844\,R_1^6\,R_2^2 \nonumber \\
& + 22898\,R_1^4\,R_2^4 - 
     11844\,R_1^2\,R_2^6 + 875\,R_2^8 \big) \,r^6 \\
& - 2\,\big( 2275\,R_1^8 + 13552\,R_1^6\,R_2^2 - 38502\,R_1^4\,R_2^4 \nonumber \\
& + 13552\,R_1^2\,R_2^6 + 2275\,R_2^8 \big) \,r^8 \nonumber \\
& + 28\,\left( R_1^2 + R_2^2 \right) \,
   \left( 325\,R_1^4 - 838\,R_1^2\,R_2^2 + 325\,R_2^4 \right) \,r^{10} \nonumber \\
& - 168\,\left( 25\,R_1^4 - 77\,R_1^2\,R_2^2 + 25\,R_2^4 \right) \,r^{12} \nonumber \\
& - 420\,\left( R_1^2 + R_2^2 \right) \,r^{14} + 525\,r^{16} \Big]\ . \nonumber 
\end{align}
\end{appendix}


\newpage

\begin{figure}
  \begin{center}
    \epsfig{file=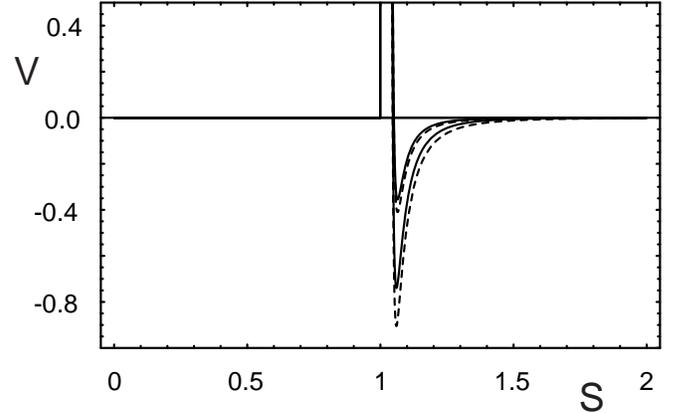,width=\columnwidth}
  \end{center}
  \caption{Interaction potential for multi-walled nanoparticles: 
    Full potential (solid) and Derjaguin approximation (dashed) for
    $\eta = 2 R / \sigma = 10$ and $\nu = h / \sigma = 0.8$ (lower
    curves) and $0.4$ (upper curves). All potentials are scaled with
    the same factor as in \fig{fig:ScaledPotentials}. The Derjaguin
    approximation improves with increasing wall thickness $\nu$.}
  \label{fig:FullPotentials}
\end{figure}

\end{document}